\documentclass[osajnl,twocolumn,showpacs,superscriptaddress,10pt]{revtex4-1}

\usepackage{amsmath,amssymb,graphicx}

\begin{document}

\title{Photo-acoustic sensor for detection of oil contamination in compressed air systems}

\author{Mikael Lassen}\email{Corresponding author: ml@dfm.dk}
\affiliation{Danish Fundamental Metrology, Matematiktorvet 307, DK-2800 Kgs. Lyngby, Denmark}

\author{David Baslev Harder}
\affiliation{Danish Fundamental Metrology, Matematiktorvet 307, DK-2800 Kgs. Lyngby, Denmark}

\author{ Anders Brusch}
\affiliation{Danish Fundamental Metrology, Matematiktorvet 307, DK-2800 Kgs. Lyngby, Denmark}

\author{Ole Stender Nielsen}
\affiliation{Danish Fundamental Metrology, Matematiktorvet 307, DK-2800 Kgs. Lyngby, Denmark}

\author{ Dita Heikens}
\affiliation{VSL - The Dutch Metrology Institute, Thijsseweg 11, 2629 JA Delft, The Netherlands}

\author{Stefan Persijn}
\affiliation{VSL - The Dutch Metrology Institute, Thijsseweg 11, 2629 JA Delft, The Netherlands}

\author{Jan C. Petersen}
\affiliation{Danish Fundamental Metrology, Matematiktorvet 307, DK-2800 Kgs. Lyngby, Denmark}


\begin{abstract}
We demonstrate an online (in-situ) sensor for continuous detection of oil contamination in compressed air systems complying with the ISO-8573 standard. The sensor is based on the photo-acoustic (PA) effect.  The online and real-time PA sensor system has the potential to benefit a wide range of users that require high purity compressed air. Among these are hospitals, pharmaceutical industries, electronics manufacturers, and clean room facilities. The sensor was tested for sensitivity, repeatability, robustness to molecular cross-interference, and stability of calibration. Explicit measurements of hexane (C${_6}$H$_{14}$) and decane (C${_{10}}$H$_{22}$) vapors via excitation of molecular C-H vibrations at approx. 2950 cm$^{-1}$ (3.38 $\mu$m) were conducted with a custom made interband cascade laser (ICL). For the decane measurements a (1 $\sigma$) standard deviation (STD) of 0.3 ppb was demonstrated, which corresponds to a normalized noise equivalent absorption (NNEA) coefficient for the prototype PA sensor of 2.8$\times 10^{-9}$ W cm$^{-1}$ Hz$^{1/2}$.
\end{abstract}

\maketitle

\section{Introduction}

Compressed air is an essential asset for many industries today. It is safe and relatively inexpensive to operate and very reliable \cite{Radgen2001,Hirzel2010}. However, compressed air is susceptible to oil contaminants, which even in minuscule quantities can be disastrous for a number of industrial applications and be health threatening for humans. In 1991 the International Standards Organization (ISO) established the ISO-8573-1 standard on purity of compressed air \cite{ISO}. This was done in order to govern compressed air system component selection, design and measurement. According to this standard all hydrocarbons with 6 or more carbon atoms per molecule are considered as "oil".  The following organic compound classes (hydrocarbons) are therefore all "oils": solvents (e.g. toluene, hexane, decane,... ), VOCs (Volatile Organic Compounds), adhesives, thread and surface sealants, fragrances (air fresheners, perfumes, etc.), detergents/cleaning agents. When oil is present in pipelines of the compressed air system, it has always one of the three forms: Aerosols, wall flow and/or vapors (oil mist). Aerosols are partially removed by coalescing filters and appear as condensate. Wall flow either appears in condensate or travels along the walls of the pipeline to the end user, again partially removed by coalescing filters.  Vapors and oil mist are not removed by coalescing filters and can therefore be a huge problem. The lack of any reliable, highly sensitive, online sensor system has forced critical industries to rely on manual sampling and subsequent laboratory analysis, which is labour intensive, inefficient, and cannot guarantee conformance of compressed air systems to the mandatory or industry adopted regulatory norms. The main characteristics of the sensor should therefore be its capability to do online measurements of "oil" contaminant concentration at 8 ppb or lower in order to meet the sensitivity requirements of the ISO-8573 standards for compressed air purity levels of Class 1 or better. Two kinds of oil are used for air compressors today: synthetic and mineral oils. According to a 2007 publication by Colyer \cite{Colyer2007}, classical hydrocarbons had a  market share of about 80$\%$, however polyglycol lubricants are claimed to have numerous benefits for use in compressors and are heavily promoted by compressed air companies.

\begin{figure}[htbp]
\centering\includegraphics[width=\linewidth]{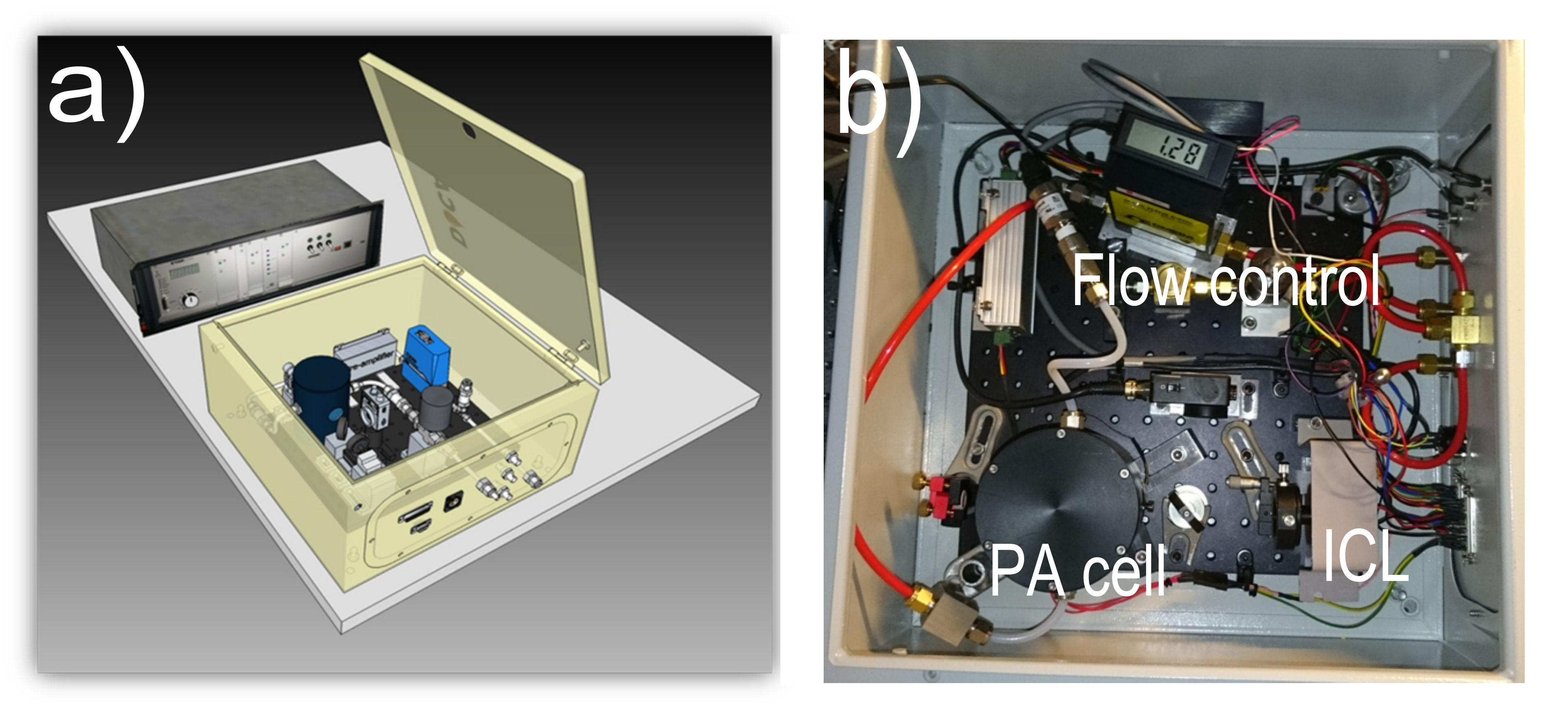}
\caption{a) 3D design drawing of the prototype sensor. The sensor platform consists of three units: the sampling system (not shown), the optical spectroscopy unit, and the electronics/software. The front box is the PA sensor head, while the rear box is the system control unit. The system control unit consist of a combined field-programmable gate array (FPGA) and microprocessors. Through software the system control unit provides control and monitoring of the vital parts in the PA sensor. b) Picture of the inside of the PA sensor head. \label{fig1} }
\end{figure}

We demonstrate an oil contamination sensor for online monitoring of compressed air. The sensor is based on the photo-acoustic (PA) effect \cite{Lopes2010,BEllPAS1881,Rosencwaig1980Book,Harren2000,Tam1986}. Photo-acoustic spectroscopy (PAS) is an established technique for environmental, industrial, and biological monitoring and imaging, due to its ease of use, compactness and its capability of allowing trace gas measurements at the sub-parts per billion (ppb) level \cite{Chen2014,Wang2012,Spagnolo2012,Lassen2015,Peltola2015,Szabo2013,Patel2008,Lassen2016}. The design and layout of the prototype PA sensor is depicted in Fig.~\ref{fig1}. \cite{patentDOCA}.

The PA system consists of a sample collector piece that is installed into the compressed air system, using the so-called isokinetic sampling method, not shown in the figure, and two units; a PA sensor module and a driver module. The PA technique is based on the detection of sound waves that are generated due to molecular absorption of modulated optical radiation. The generated PA signal is proportional to the density of molecules, which makes the PA technique able to measure the absorption directly, rather than relying on having to calculate it from the transmission of the radiation. The PAS technique is however not an absolute technique and calibration is required against a known sample (known concentrations of the gas/oil).

Explicit measurements of hexane (C${_6}$H$_{14}$) and decane (C$_{10}$H$_{22}$) vapors via excitation of molecular C-H vibrations at approx. 2950 cm$^{-1}$ (3.38 $\mu$m) were conducted with an interband cascade laser (ICL) \cite{Yaws1999}. The PA sensor has an oil concentration sensitivity lower than 1 ppb with traceability to certified reference gases (verified by VSL, The Dutch Metrology Institute), applicable for Class 1 detection (the most stringent classification according to ISO 8573). A linear dependency between the spectroscopic signal and the hexane and decane amount of substance fractions was observed, confirming the proper functioning of the PA sensor for all classes of oil contamination. The interface between the compressed air system and the online PA sensor is critical as it warrants that representative samples of the compressed air are taken. The sample system should be able to function up to a pressure of 16 bar and be able to stand pressure pulses up to 32 bar (<1 second). The design is therefore robust and heavy designed in order to withstand high pressures. Continuously air samples are taken using the so-called isokinetic sampling method. The sensor is flowed continuously with 1.7 L/min for fast sampling of the compressed air. The sampling probe is placed at the center of the pipes of the compressed air system. Due to the particular choice of tube lengths and diameters, the speed of air in the compressed air system is the same as the speed of air in the probe. More details on the isokinetic sampling can be found in the ISO 8573 standard \cite{ISO}.

The sensor was tested for the following parameters: Sensitivity (ppb level), accuracy and standard deviation (STD) of the measurements, long term stability, response time, reproducibility and bias/background signals, self-cleaning methods and flow noise immunity.  In this contribution we present data for the sensitivity, response time and reproducibility of the PA sensor and the stability over time. The ability to detect oil contaminants online, will significantly enhance the capabilities of manufacturers and users to guarantee the quality of their products and eliminate a number of risks and civil liabilities that are associated with non-conformance.

\section{The photo-acoustic effect}

The sensor is based on the photo-acoustic effect \cite{Rosencwaig1980Book,BEllPAS1881}. The PA technique detects sound waves that are generated due to absorption of modulated optical radiation in various molecular species. In a gas the sound waves are generated due to local heating via molecular collisions and de-excitation. Absorption of laser radiation excites the molecules in the PA cell. The added energy is via collisional processes converted to local heating and de-excitation of the molecules in the PA cell thus generating sound waves. A pressure sensitive device (e.g. microphone, tuning fork or cantilever) is used to monitor these modulated sound waves. The magnitude of the measured PA signal in volts is given by:
\begin{equation}
 S_{PA} = S_m P F \alpha,
 \label{eq.PAsignal}
\end{equation}
where $P$ is the power of the incident radiation, $\alpha$ is the absorption coefficient, which depends on the total number of molecules per cm$^3$ and the absorption cross section, $S_m$ is the sensitivity of the microphone and $F$ is the cell-specific constant, which depends on the geometry of acoustic cell and the quality factor $Q$ of the acoustic resonance \cite{Rosencwaig1980Book}. Ideally a highly sensitive PA sensor should only amplify the sound waves and reject acoustic and electrical noise as well as in-phase background absorption signals from other materials in the cell (walls and windows).

\section{Experimental setup}

\begin{figure}[htbp]
\centering\includegraphics[width=\linewidth]{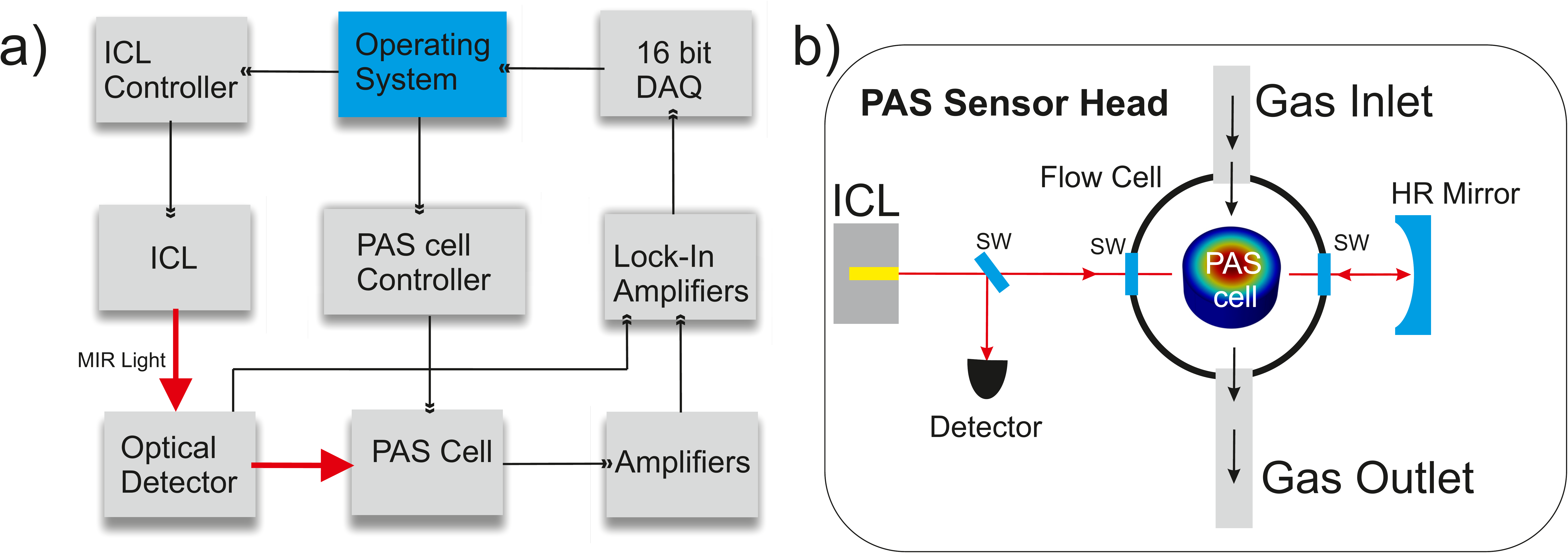}
\caption{a) Block diagram of the main parts of the setup. The PAS cell controller regulates the temperature, the flow and the pressure inside the cell. The ICL controller controls the temperature and the modulation of the ICL current. b) Shows the optical parts of the sensor head and inner PAS cell. A 3.38 $\mu$m laser beam from the ICL (Interband Cascade Laser) is aligned to make a double pass through the cell in order to enhance the interaction with the oil molecules. Mount for the ICL; PAS cell; back coupling mirror; SW: Silicon Window. The center of the PAS cell shows a COMSOL simulation of the acoustic resonance mode at 6.5 kHz. Two 30 mm in diameter plates constitute the acoustic resonator, and are the main components of the PAS cell. \label{fig2}}
\end{figure}

\begin{figure}[htbp]
\centering\includegraphics[width=\linewidth]{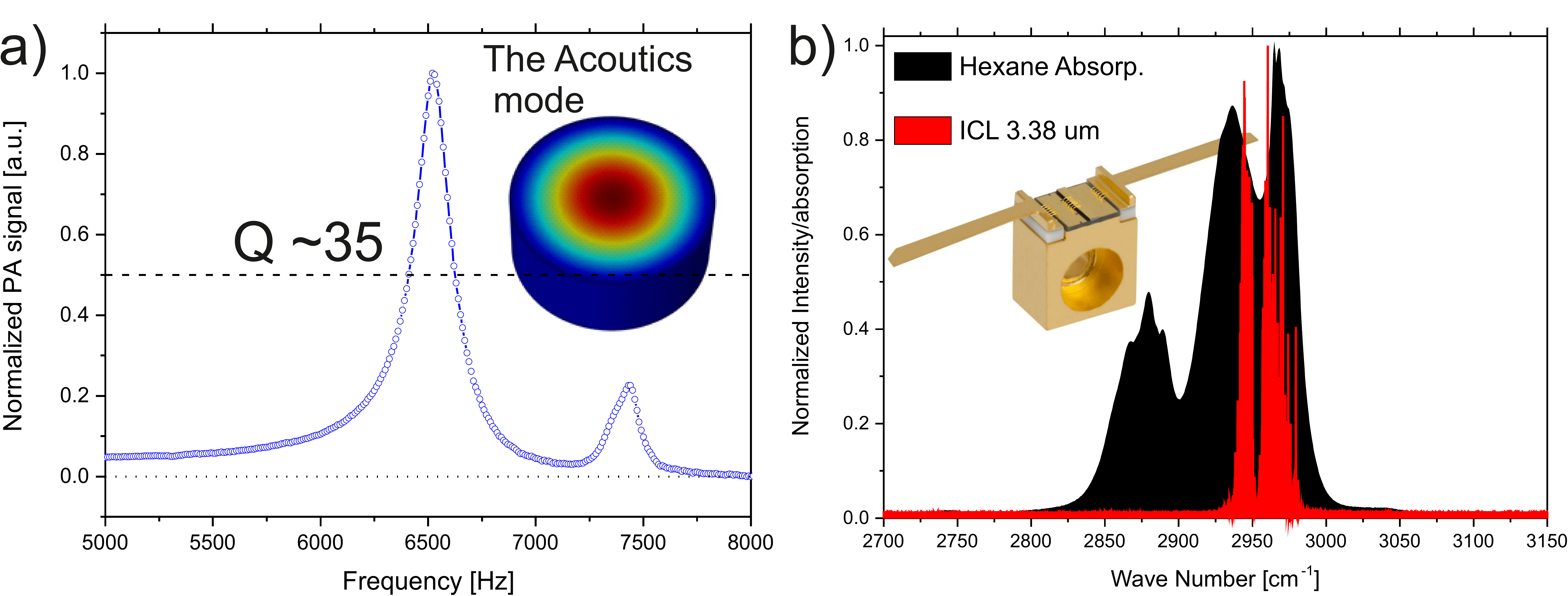}
\caption{a) Shows the acoustic resonance at approximately 6.5 kHz and the associated drum skin mode. The visualization of the drum skin mode was simulated using COMSOL. b) The normalized spectrum of the ICL (red curve) centered at 3.38 $\mu$m. Plotted together with a Hexane absorption spectrum (black curve).   \label{fig3}}
\end{figure}

The block diagram of the main parts of the setup and the schematics of the sensor head are shown in Fig.~\ref{fig2}. A typical setup for PAS involves an amplitude modulated light source and a resonant absorption cell with microphones, where the PA signal is enhanced by the acoustic resonances. The highest conversion efficiency of laser power to PA signal is obtained when the laser spectrum is narrower than the absorption feature, since laser power at frequencies outside the absorption feature will not contribute to the PA signal.  The C-H stretch band of the oils to be monitored are located at approximately 3.38 $\mu$m and is shown for hexane in Fig.~\ref{fig3}(b). This band is very broad and spectrally dense, thus high resolution spectroscopy probing individual ro-vibrational absorption lines in the oil molecules is not required. Therefore a broadband custom made ICL is used. The ICL spectrum is shown in Fig.~\ref{fig3}(b) together with the hexane absorption spectrum. Additional requirements to consider are the power level required for the PAS measurements, compactness, form factor and price. The laser is temperature controlled with a Peltier element and kept at 20$^\circ$C (+/- 0.1$^\circ$C). If the temperature of the ICL is not kept constant the laser spectrum shifts in frequency. This will change the overlap between the laser spectrum and the oil absorption spectrum resulting in a changed PA signal and thus in a change in the evaluated oil concentration level. The modulation of the light intensity is controlled by modulating the current, where a square-wave is fed into the laser controller. The duty cycle used is always 50/50. The laser controller can be operated with frequencies from 100 Hz to 20 kHz. The output beam from the ICL is collimated with a molded IR aspheric lens. The lens is AR coated in the wavelength region 3 - 5 $\mu$m. After the collimating lens a beamsplitter is inserted to tap approximately 0.5$\%$ of the laser light onto a MIR photo detector (PbSe, 1.5-4.8 $\mu$m, AC-Coupled Amplifier) for optical power measurement and normalization of the PA signal. The optical transmission through the cell is approximately 97$\%$ at 3.38 $\mu$m and the absorption coefficient is approximately 10$^{-3}$ cm$^{-1}$ for each cell window. The 3.38 $\mu$m ICL can deliver over 60 mW of optical power. The PA signal can be enhanced by optical multi-pass techniques resulting in an increase of the sensitivity of the PA spectrometer due to the increased light absorption path length from multiple reflection. Various multi-pass and single-pass configurations have so far been exploited for PAS configurations, such as ring cells, cavity based cells and transverse square cells  \cite{Nägele2000,Rey2005,Miklos2006,Saarela2010,Manninen2012,lassen2014}. Here we use a double pass configuration with a highly reflection mirror (on the order of 99$\%$ and a radius of curvature of 500 mm) to reflect the beam back into the PA cell as illustrated in Fig.~\ref{fig2}(b). The double pass configuration is used in order to enhance the interaction with the oil molecules, thereby enhancing the PA signal and the sensitivity.

The acoustic resonator of the PAS cell has an open cell configuration and is made by two circular plates with a diameter of 30 mm and separated by a distance of 10 mm.  The acoustic response of the PA cell is shown in Fig.~\ref{fig3}(a). In order to have a continuous flow of air through the PAS cell (up to 2 l/min), input nozzles and output nozzles have been attached on opposite sides of the flow cell. The dimensions of the nozzles are 10 mm in diameter and the flow is kept stable using massflow controllers. The acoustic resonator is covered by an outer cell with an 80 mm diameter. It acts as the buffer zone for the acoustic flow noise and the acoustic signal generated by window absorption. The design of the buffer flow zone has been both theoretically and experimentally investigated. The theoretical investigation included the use of the OpenFOAM software, however the simulations are not shown here. The aim of the analysis was to investigate and optimize the influence of the cell geometry and flow rate on the gas distribution within the cell and how flow noise is affecting the acoustics resonator. The PA cell is heated to 65$^\circ$C ($\pm$ 0.1$^\circ$C).  The temperature control of the PA is very important since the exact resonance frequency is a function of temperature and shifts in temperature will add an uncertainty to the PA signal.

The microphone is positioned in the middle of the plates, thus at the maximum acoustic amplitude. The signal from the microphone were amplified with a homebuild low noise amplifier with variable gain and filtered with a 7 kHz bandpass filter (3-10 kHz) before further signal processing. All data was processed using a lock-in amplifier. The ICL modulation is controlled by a signal generator, which also acts as the local oscillator for the lock-in amplifier. The data from the lock-in amplifier is collected with a 250 kS/s data acquisition (DAQ) card with 16 bit resolution.

\section{Results using certified oil samples prepared at VSL}

\begin{figure}[htbp]
\centering\includegraphics[width=11cm]{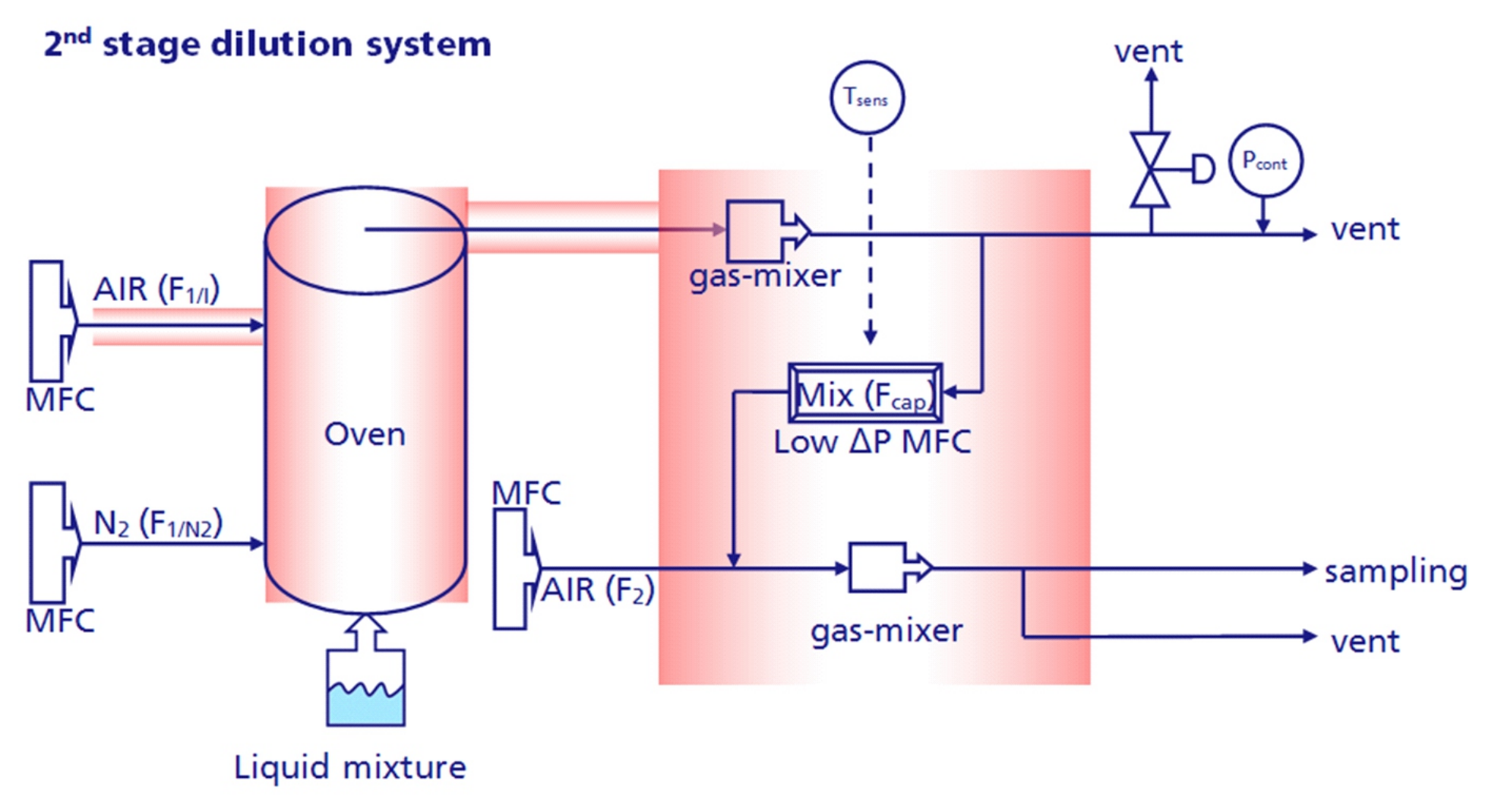}
\caption{The two stage dilution system set up used during tests at VSL. The VSL measurements are conducted by measuring the mass loss of the liquid mixture over time with a sensitive balance. Information about the mixing ratios for the various dilution steps determines the concentration.\label{fig4}}
\end{figure}

The PAS technique is not an absolute technique and requires calibration using a certified gas reference sample. The experiments were performed by excitation of molecular C-H ro-vibrational modes of hexane and decane. The samples were prepared using the dynamic generation system at VSL, the Dutch Metrology Institute. Part of this system is shown in Fig.~\ref{fig4}. The dynamic system is a two-step dilution system. A small amount of substance (in liquid phase) is forced via a small capillary into an oven, which is heated and flowed with known amounts of nitrogen and pure air.
The next dilution step further decreases the concentration by dilution with a known amount of pure air. The mass loss is measured every 60 seconds. Knowledge of the mass loss of the substance together with the various flow rates allows the hexane and decane concentration to be calculated with high precision. In the following data all concentrations and associated uncertainties for the hexane and decane are determined by the mass loss of the oil substance. Note that all measurements with the PA sensor have been tested experimentally with 1.7 l/min of flow and the PA sensor cell pressure was maintained at approximately 1 bar for all experiments.

\begin{figure}[htbp]
\centering\includegraphics[width=\linewidth]{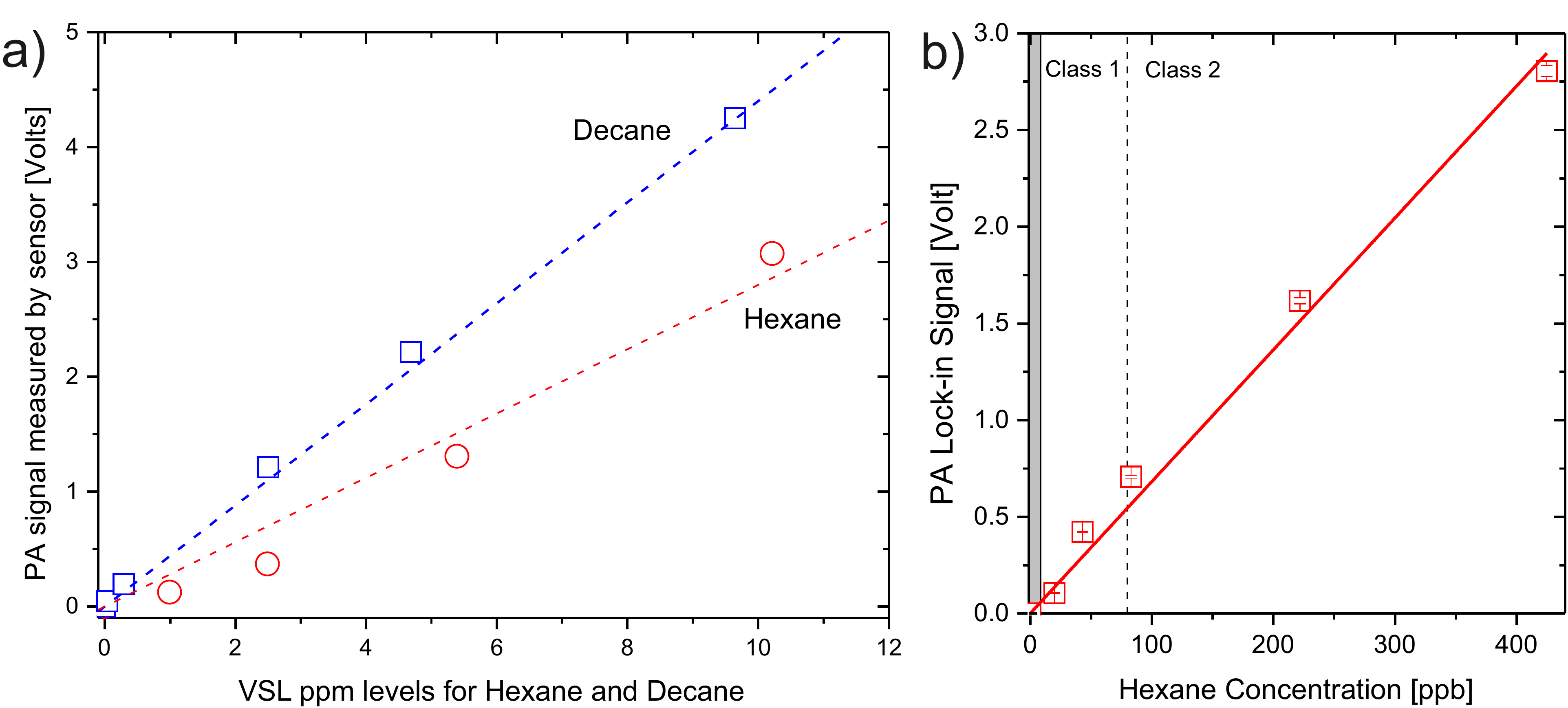}
\caption{The sensors linear dependency between the lock-in amplifier signal and the hexane and decane  amount of substances (concentrations). a) ISO class 3 and ISO class 4 measurements of Decane and Hexane. b) ISO class 1 (below 80 ppb) and ISO class 2 (below 800 ppb) measurements of decane and hexane. Note that ppm: $\mu$mol/mol and ppb: nmol/mol. \label{fig5}}
\end{figure}

Figure~\ref{fig5}(a) demonstrates the linear dependency between the lock-in amplifier signal and the hexane and decane concentrations with an uncertainty of less than 3$\%$ for each measurement. The fitted line for the decane measurements has the best linear response. This is probably due to a small drift in the alignement and stabilization of the gas sample in the PA sensor. Since the decane concentrations were measured within a few hours while the hexane measurements were measured during 2 days without realignment of the optical system. Fig.~\ref{fig5}(b) shows measurements of hexane measured within a few hours and it is observed that the linearity of the PA sensor is better. These results also stress the importance for calibrating and recalibrating of the PA sensor in order to achieve trusted values for the concentrations.  From  Fig.~\ref{fig5}(a) it can be seen that the decane signal is approximately 1.5 times larger than the hexane response which is in good agreement with the normalized absorption as a function of carbon number \cite{PNNLdata}. The data shown in Fig.~\ref{fig5} is processed with a lock-in amplifier with an integration time of 3 seconds.

\begin{figure}[htbp]
\centering\includegraphics[width=\linewidth]{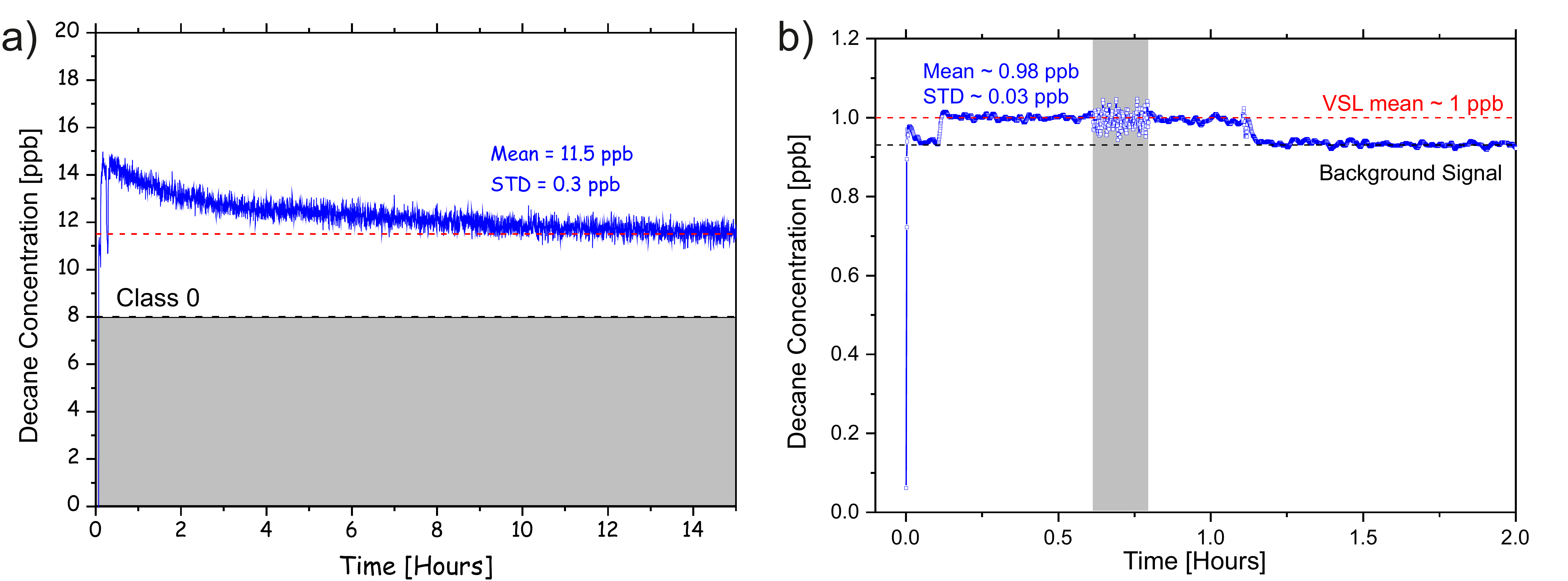}
\caption{a) Concentration measurements of decane (C$_{10}$H$_{22}$) as a function of time using the PA sensor. The mean concentration used was 11.5$\pm4.4$ ppb with a standard deviation of 0.3 ppb. The measurements are processed with a lock-in amplifier with an integration time of 10 seconds. b) The PA background signal level at 0.93 ppb and measurement of 1 ppb decane concentration. The integration time was 10 seconds on the lock-in amplifier, except for the data in the gray shaded area for which it was 1 second.  \label{fig6}}
\end{figure}

Figure~\ref{fig6}(a) shows the result of measurements performed on an 11.5$\pm4.4$ ppb decane certified reference sample.  The voltage signal from the PA sensor is calibrated using this value. It is clearly seen that the sensor is measuring a slightly higher concentration of approximately 2 ppb for the first 7 hours. This is attributed to oil condensation in the diluting system, since oils are "sticky" molecules and have tendency to adhere to surfaces, which slowly evaporate and therefore gives a slightly higher concentration going into the sensor than what is being prepared. However, after 7 hours of flushing the system with an 11.5 ppb decane concentration in air, it finally finds a stable level. The measurements after 8 hours have (1$\sigma$) STD of 0.3 ppb. Decane has a maximum absorption of 5.5$\times$10$^{-5}$ cm$^{-1}$ for 1 $\mu$mol/mol at 3.38 $\mu$m, thus the maximum normalized noise equivalent absorption coefficient is 2.8 $\times 10^{-9}$ W cm$^{-1}$ Hz$^{1/2}$. This makes the PA sensor comparable with state of the art PA sensors \cite{Patimisco2014,Peltola2015,Sampaolo2016}.  The data shown in Fig.~\ref{fig5} and Fig.~\ref{fig6} demonstrates the potential of the PA sensor for measuring all five ISO classes of oil contamination in compressed air.

Figure~\ref{fig6}(b) shows the measurement of the background signal and a class 0 decane concentration. The dilution stage measured a mean concentration of 1$\pm 2.53$ ppb over the 2 hours of measurements. This value is used to normalize the data seen in Fig.~\ref{fig6}(b). The background signal due to unwanted absorptions in the PA cell (windows, acoustic plates, ...) is approximately 0.93 ppb. This value determines the absolute sensitivity of the PA cell. The graph shows measurements of a 1 ppb decane concentration with a S/N (Signal to background) of 1.1. This means that the S/N for a class 1 detection would be better than 11 and this demonstrates that the prototype PA sensor has the needed sensitivity to detected class 1 or better concentrations.  All data shown in Fig.~\ref{fig5} are processed with a lock-in amplifier with an integration time of 10 seconds, except the gray shaded area which were processed with a lock-in amplifier with a integration time of 1 second. Note that using heavier oils e.g. going from hexane to decane as demonstrated, improves the sensitivity of the PA sensor we would therefore expect for typically oils used for compressors (with up to ~C$_{80}$) that the sensitivity would be one order of magnitudes higher \cite{PNNLdata}.

\begin{figure}[htbp]
\centering\includegraphics[width=\linewidth]{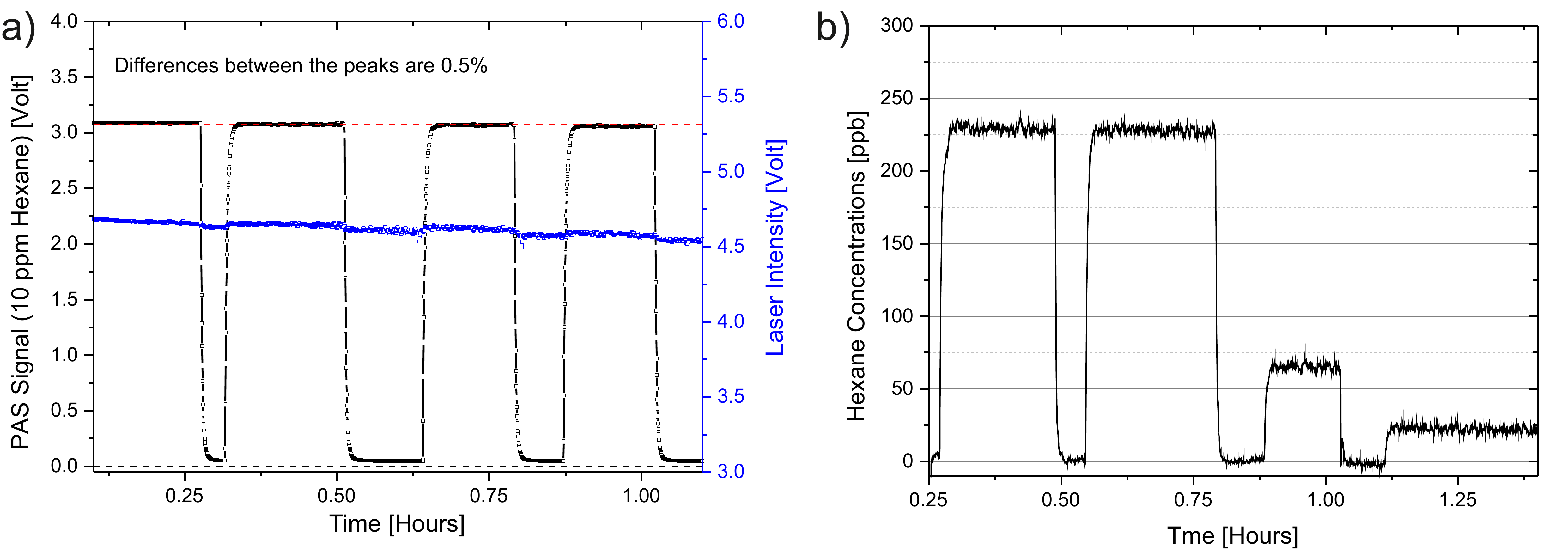}
\caption{Test of reproduciblility and response time of the sensor. a) The sensor is flowed with 10 ppm of hexane followed by a flow with pure air. The blue curve shows the laser intensity over the measurement time. b) The sensor is flowed with 222 ppb of hexane followed by a flow with pure air. Then flowed with 65 ppb and 24 ppb of hexane followed by a flow with pure air. \label{fig7}}
\end{figure}

In order to test the reproducibility and the response time of the sensor the sensor was flowed with two different concentrations of hexane,  10 ppm and 222 ppb, respectively, followed by a flow with pure air. Figure~\ref{fig7}(a) shows that the reproducibility of the sensor system signal when using a very high concentration of oil (10 ppm). The reproducibility  is within 0.5$\%$ peak to peak. The response time of the sensor is fairly quick and that level stabilizes within 30 seconds. Figure~\ref{fig7}(b) shows that the reproducibility when the system is flowed with 222 ppb of hexane of the PA sensor signal is within 1$\%$ from peak to peak and with a class 1 sensitivity (STD < 8 ppb). The response time is around 15 seconds for reaching the full level of 222 ppb again. However in both for the high concentration and the low concentration the response time depends on the lock-in amplifiers integration time (in these measurements the integration time of the lock-in amplifier was 3 seconds). From this we conclude that the time response of the sensor will not be the limiting factor for notifying the customer or end user of an oil breach in the air compressor filters.

\section{Conclusion}

In conclusion the prototype PA sensor provides continuous online (in-situ) measurements of oil contaminations in compressed air systems. The PA sensor provide a unique sensitivity that allows detection of less than 1 ppb contaminations levels of all lubricates and combinations of oils. The prototype PA sensor with flow buffer zone has been tested experimentally with 1.7 l/min flow and mean decane concentration measurements down to 1 ppb were demonstrated. It was found that after 8 hours of flushing the PA sensor with a decane concentration of 11.5 ppb in air, a (1$\sigma$) STD was of 0.3 ppb was obtained, and thus a normalized noise equivalent absorption coefficient of 3.1 $\times 10^{-9}$ W cm$^{-1}$ Hz$^{1/2}$ was demonstrated. Further the measurements demonstrate that the PA sensor fulfills many of its requirements, namely the sensitivity, reproducibility, flow immunity, response time, linearity of the PA signal to oil any oil concentration within the ISO-8573 classes and repeatability of the background bias signal (no contamination). To summarize the PA sensors specifications are: sensitivity: 0.3 ppb (class 0: <8 ppb), reproducibility within $1\%$, time response less than 15 seconds, flow immunity up 2 l/min flow without reducing the sensitivity.

A recently started Eurostars project titled PASOCA (Photo-Acoustic Sensor for Oil detection in Compressed Air) will take the developed prototype PA sensor without impairing it to a certified commercial product. We therefore believe that with further miniaturization and proper mechanical design the PA sensor technology developed here will be ready for mass production. We foresee that the developed oil contamination PA sensor will benefit a large category of industries that require high purity compressed air including hospitals, pharmaceutical industries, electronics, clean rooms and many other industries which are in need of pure compressed air.

\section*{Funding and Acknowledgments}

We acknowledge the financial support from EUREKA (Eurostars program: E10132 - PASOCA) and the Danish Agency for Science, Technology and Innovation. The European Unions Seventh Framework Programme (FP7) managed by REA 8211 under Grant Agreement N. 286106. We would like to thank Poul Jessen and S{\o}ren Laungaard from PAJ Group (paj@paj.dk) for fruitful discussions and collaboration.

\end{document}